\documentclass[12pt]{article}
\usepackage{amsmath,amscd,amsfonts,eucal,latexsym,amssymb,mathrsfs}
\usepackage{amsxtra, amsthm, calc}
\usepackage{graphicx}
\usepackage[dvipsnames]{xcolor}
\usepackage{{setspace}}
\newlength{\dinwidth}
\newlength{\dinmargin}
\newsymbol\rest 1316         

\newsymbol\bt 1202  

\newcommand{\hide}[1]{} 

\newcommand{\Pport}{\mathcal{P}_+^\uparrow}

\newcounter{propcount}

\newlength{\maxlabelwidth}

\newcommand{\aaa}{\mbox{\footnotesize{Alice}}}
\newcommand{\bbb}{\mbox{\footnotesize{Bob}}}
\newcommand{\ccc}{\mbox{\footnotesize{Charlie}}}
\begin{document}
\noindent
\large
${}$ 
\begin{center}
{\Large Superluminal local operations in quantum field theory: \\ A ping-pong ball test}
\\[16pt]
{\bf Albert Much} \ \  and \ \ {\bf Rainer Verch}
\\[16pt]
Institut f\"ur Theoretische Physik, Universit\"at Leipzig, Germany\\[6pt]
{\small much{@}itp.uni-leipzig.de, rainer.verch{@}uni-leipzig.de}\\[6pt]
{\small Revised version, February 2024}
\end{center}
${}$
\\ \\
{\bf Abstract} \ \
It is known that in quantum field theory, localized operations, e.g.\
given by unitary operators in local observable algebras, may lead to non-causal, or superluminal, state changes within their localization region. In this article, it is shown that both in quantum field theory as well as in classical relativistic field theory, there are localized operations which correspond to ``instantaneous'' spatial rotations (leaving the localization region invariant) leading to 
superluminal effects within the localization region. This shows that ``impossible measurement scenarios'' which have been investigated in the literature, and which 
rely on the presence of localized operations that feature superluminal effects within their localization region, do not only occur in quantum field theory, but also in classical field theory.

\section{Introduction}
There are some scenarios, usually set within the framework of special relativity, in 
which it is argued that superluminal effects are related to effects that 
are akin to travelling backward in time -- see, e.g., Sec.\ 4.3 in \cite{Sho}; see, however, also \cite{AMNS} and references given there for more critical considerations
on this issue. Taken for granted that a compelling connection between superluminal 
effects and time travel can be established, our present contribution fits into the 
theme of this volume.

Recently, some attention has been given to the circumstance that there are ``local
unitary operations'' in relativistic quantum field theory which however can act in 
a ``superluminal'' fashion within their localization region \cite{Jubb,AJu,BorJubKel}. This has, in fact,
been observed much earlier by Sorkin \cite{Sorkin}, who employed it to argue that 
relativistic quantum field theory was lacking a well-defined approach to measurement comparable to the theory of measurement in non-relativistic quantum mechanics. 
To illustrate his point, he considers three spacetime regions, $O_{\aaa}$, $O_{\bbb}$ and 
$O_{\ccc}$ wherein and during which the observers Alice, Bob and Charlie can carry out
operations and measurements on a state of a quantum field they jointly have access to.
 The spacetime regions $O_{\aaa}$ and $O_{\ccc}$ are causally separated, but there is 
causal contact of $O_{\bbb}$ with both $O_{\aaa}$ and $O_{\ccc}$. (See Figure 1 below for an illustration. In some publications, like \cite{Jubb} and \cite{AJu}, the 
roles of $O_{\bbb}$ and $O_{\ccc}$ are interchanged; our labelling coincides with that in \cite{Sorkin} and \cite{BosFewRup}.) Sorkin then argues 
that there are certain combinations of unitary operations carried out by Alice in 
$O_{\aaa}$ and by Bob in $O_{\bbb}$ so that, if Charlie measures the resulting 
state in $O_{\ccc}$, it can be determined whether Alice has carried out her operation,
despite the fact that $O_{\aaa}$ and $O_{\ccc}$ are not in causal contact. Notice that,
if Bob does not carry out any operation, then Charlie cannot decide by measurements in
$O_{\ccc}$ if Alice has carried out a unitary operation in $O_{\aaa}$.

We will describe the set-up of \cite{Sorkin} in more detail below (in a version given 
by \cite{BosFewRup}), and will 
show that indeed there are local unitary operations with the properties just described.
In response to the apparent superluminal transformations of states by local unitary operations and the ensueing
difficulties regarding measurement in relativistic quantum field theory according to Sorkin, it has been shown in
\cite{BosFewRup} that these difficulties do not occur in a recently proposed, covariant approach to local measurement in quantum field theory \cite{FVqflm} (see also \cite{FewQFM,FVenc,FraPap} for additional discussion). In the present paper, we wish to 
point out that certain ``superluminal localized operations'' aren't specific to relativistic quantum
field theory (and therefore, relativistic quantum field theory isn't suffering from 
any particular conceptual defect or inconsistency), but they appear also in classical relativistic theories. For instance, they are related to (local) symmetries that a theory, quantum or classical, may 
possess, but which cannot be performed ``instantaneously'' as they violate the principles of special or 
general relativity both on kinematical as well as dynamical grounds. These are occasionally (but perhaps not systematically) referred to as ``passive'' symmetry transformations.  Therefore, the present paper provides a ``ping-pong ball test'' as regards the occurence of ``superluminal localized operations''. Said test is a concept which, according to the present authors' knowledge, goes back to Reinhard Werner \cite{Wer-Inv}, and we paraphrase it here as follows: {\it When someone presents a paradox as being rooted in quantum physics, replace the term `quantum mechanical particle' by `ping-pong ball' everywhere. If the paradox persists, it is unrelated to quantum physics.}\footnote{In \cite{Wer-Inv}, the ping-pong ball test is more specifically related to 
Bell's inequalities, and its wording is verbatim as follows: {\it 
Take an author’s explanation of Bell’s inequalities, and substitute ``ping-pong balls'' for every quantum particle. Then if whatever the
author is selling as paradoxical, remains true, he/she hasn’t understood a thing.}}
As a matter of fact, application of the ping-pong ball test in other scenarios 
claiming that acausal effects may occur by means of quantum physics has already proven useful \cite{TolVer-2}. 

The present article is structured as follows. In Sec.\ 2, we summarize properties
of relativistic quantum field theories on $1+3$-dimensional Minkowski spacetime in the operator-algebraic framework. The assumption of the ``split property'' implies that global symmetries, such 
as space rotations, have unitary implementers in the local operator algebras. 
In Sec.\ 3, we revisit the ``impossible measurements scenario'' presented in 
\cite{Sorkin}, and we show how it can be realized by ``instantaneous space rotations'' which have unitary implementing operators contained in local algebras. The quantized Klein-Gordon field is used as a special, simple example. We show in Sec.\ 4 that in a recent proposal for an algebraic description of classical field theory in terms of local Poisson algebras, there are also local symmetries corresponding to ``instantaneous space rotations'', thus the considerations 
leading to the ``impossible measurements scenario'' apply for classical field theory as well. We discuss the conclusions that can be drawn from these results in Sec.\ 5.

\section{Algebraic quantum field theory setting}

\setcounter{equation}{0}

We start by considering relativistic quantum field theory on 1+3 dimensional Minkowski spacetime (represented as $\mathbb{R}^{1+3}$) in vacuum representation. This is mainly 
for convenience; generalizations of the arguments given below to  the case of more general (globally hyperbolic) spacetimes, or spacetime dimensions $\ge$ 1 + 2, are not difficult. 

Thus, the standard assumptions are made (cf.\ \cite{HK,Haag}): There is a Hilbert space $\mathcal{H}$ on which 
a continuous representation $U_L$, $L \in \mathcal{P}_+^\uparrow$, of the proper, orthochronous Poincar\'e group operates; there is a (up to phase) unique unit vector $\Omega \in \mathcal{H}$ which
doesn't change under the Poincar\'e transformations, i.e.\ $U_L \Omega = \Omega$. If the 
translations in $\mathcal{P}_+^\uparrow$ are denoted as $a \in \mathbb{R}^4$, and their
unitary representers as $U_a$, then for any future-directed, timelike
unit vector $e$, the unitary group $t \mapsto U_{te}$ $(t \in \mathbb{R})$ is assumed to
have a selfadjoint generator with non-negative spectrum: This is the {\it spectrum condition}.

Moreover, it is assumed that there is a family ${\sf A}(O)$  of von Neumann subalgebras of ${\sf B}(\mathcal{H})$ indexed by the open, relatively 
compact subsets $O$ of $\mathbb{R}^{1+3}$, subject to the conditions of {\it isotony}: $O_1 \subset O_2$ $\Rightarrow$ ${\sf A}(O_1) \subset {\sf A}(O_2)$, and {\it locality}: ${\sf A}(O_2) \subset {\sf A}(O_1)'$ if $O_2 \subset O_1^\perp$. Here, ${\sf A}(O_1)' = \{ C \in {\sf B}(\mathcal{H}): CA = AC \ \text{for all} \ A \in {\sf A}(O_1)\}$ is the commutant algebra (or simply {\it commutant}) of ${\sf A}(O_1)$ and we recall that any von Neumann algebra ${\sf A}$ in
${\sf B}(\mathcal{H})$ is characterized by the property that ${\sf A}'' = {\sf A}$. Furthermore, for any open subset $O$ of $\mathbb{R}^{1+3}$ we denote by $O^\perp$ the causal complement of $O$, i.e.\ the largest open set in $\mathbb{R}^{1+3}$ such that there is no 
pair of points $p \in O$ and $p^\perp \in O^\perp$ which can be connected by any smooth, causal curve. 

The algebra ${\sf A}(O)$ is viewed as the algebra of (in the sense of ``generated by'') the 
observables that can be measured within the spacetime region $O$. The unitary representation
of $\mathcal{P}_+^\uparrow$ acts covariantly on the collection of local observable algebras,
\begin{align}
  \alpha_L({\sf A}(O)) = {\sf A}(L(O)) \quad \text{where} \quad \alpha_L(A) = U_L A U_L^*\,.
\end{align}
It is assumed that the von Neumann algebra generated by all the ${\sf A}(O)$ coincides with ${\sf B}(\mathcal{H})$. 

We recall that for any  subset $S$ of Minkowski spacetime, the domain of dependence 
$D(S)$ is the set of all points $p$ in the spacetime such that all past-directed or all future-directed causal rays emanating from $p$ intersect $S$. In Minkowski spacetime, an open subset $O$ is 
called causally complete if it has the property $O = (O^\perp)^\perp$, which entails also that $O = D(O)$. 

In addition to the standard properties for a quantum field theory in the operator algebraic setting just stated, we will make a few additional assumptions which are known to hold e.g.\ in models of linear quantized fields. The first property is the {\it local time-slice property}:
${\sf A}(O) = {\sf A}(D(O))$ (sometimes also called {\it primitive causality}). This is demanded to hold for spacetime regions which are of the form $O = D(B) \cap N$ where $B$ is an open subset of an arbitrary Cauchy surface and $N$ is any open neighbourhood of the Cauchy surface. The second property is {\it additivity}: If $O$ is an open, relatively compact spacetime region, and $O_i$, $i \in J$, is any covering of $O$ by open, relatively compact spacetime regions for arbitrary index set $J$, then ${\sf A}(O)$ is contained in the von Neumann algebra generated by all the ${\sf A}(O_i)$. Together with the previously stated conditions, this entails the Reeh-Schlieder property of the local von Neumann algebras with respect to the vacuum vector $\Omega$, meaning that ${\sf A}(O)\Omega$ is dense in $\mathcal{H}$ if 
$O$ is any (non-void) open, relatively compact subset of $\mathbb{R}^{1+3}$. The third property, actually very relevant to our discussion, is the {\it split property}: Assume that 
$O_1$ and $O_2$ are relatively compact, causally complete open subsets of $\mathbb{R}^{1+3}$,
such that $\overline{O_1} \subset O_2$. Then there is a type I factor von Neumann algebra ${\sf N}$ such that 
\begin{align} \label{eq:split}
 {\sf A}(O_1) \subset {\sf N} \subset {\sf A}(O_2)\,.
\end{align}
We won't discuss this property here, except for remarking that the local von Neumann algebras 
${\sf A}(O)$ are typically type III, and that the type classification, roughly speaking, gives information about what kind of projection operators a von Neumann algebra possesses. 
The reader is referred to \cite{Haag,BDL,WerLPS} and references therein for considerably more discussion. One consequence, as shown in \cite{BDL}, is that global symmetries of the quantum field theory can be localized. Here, we are interested in a special case of that consequence, and we now introduce suitably adapted notation. An inertial system is assumed to be chosen, and the coordinates $(x^0,x^1,x^2,x^3)$ of $\mathbb{R}^{1+3}$ are the corresponding inertial coordinates. We consider the centered ball of radius $r >0$ at $x^0 = 0$,
\begin{align} \label{eq:ball}
 B(r) = \{ (x^0 = 0,x^1,x^2,x^3) : (x^1)^2 + (x^2)^2 + (x^3)^2 < r^2\}
\end{align}
and its domain of dependence (coinciding with its causal completion)
\begin{align} \label{eq:diamond}
 O(r) = D(B(r))\,.
\end{align}
If $R \in {\rm SO}(3)$ denotes any space rotation in the $x^0 = 0$ hyperplane around $x^j = 0$ $(j = 1,2,3)$, whereby it 
is canonically identifyable with an element in $\Pport$, then $R B(r) = B(r)$ and 
$RO(r) = O(r)$. Consequently, denoting by $U_R$ the unitary implementer of $R$, we have 
\begin{align}
 U_R {\sf A}(O(r)) U_R^* = {\sf A}(O(r))\,.
\end{align}
For any positive numbers $r_1 < r_2$, we have $\overline{O(r_1)} \subset O(r_2)$. Since the 
split property \eqref{eq:split} holds for $O_j = O(r_j)$ $(j = 1,2)$, the results of \cite{BDL} show 
that there is a continuous unitary representation $\check{U}_R$, $R \in {\rm SO}(3)$,
with the properties
\begin{align}
 \check{U}_R \in {\sf A}(O(r_2)) \quad \text{and} \quad 
 \check{U}_R A_1 \check{U}_R^* = U_R A_1 U_R^*
\end{align}
for all $A_1 \in {\sf A}(O(r_1))$ and all $R \in {\rm SO}(3)$. The $\check{U}_R$ are therefore ``localized versions'' of the unitary implementers $U_R$ of space rotations $R$.

Note that a space rotation $R$ by any finite angle acts instantaneously and therefore, with 
superluminal speed. To illustrate this to its extreme, let $R_{\pi,3}$ be a rotation by the angle $\pi$ around  $e_3$ where $e_j$ denotes the space unit vector along the 
$x^j$ coordinate axis in the $x^0 = 0$ hyperplane. Then we obtain, e.g., for 
positive numbers $s$ and $\lambda$ such that $s + \lambda < r_1$, and defining
\begin{align}
 O^{(\pm)} = D(B(s) \pm \lambda e_1)\,,
\end{align}
that $O^{(\pm)} \subset O(r_1)$ and
\begin{align}
 R_{\pi,3} (O^{(\pm)}) = O^{(\mp)}\,,
\end{align}
implying
\begin{align}
 \check{U}_{R_{\pi,3}}{\sf A}(O^{(\pm)}) \check{U}_{R_{\pi,3}}^* = 
  {\sf A}(O^{(\mp)})
\end{align}
which means that the adjoint action of $\check{U}_{R_{\pi,3}}$ rotates observables
localized in $O^{(\pm)}$ ``instantaneously'' to localization in $O^{(\mp)}$.

\section{Superluminal localized state transformations in\\ quantum field theory}

\setcounter{equation}{0}

Let us recall some further concepts which are relevant to our discussion. In what follows we will consider density matrix states for the quantum field theory described before. 
That means, if $\varrho$ is a density matrix operator on $\mathcal{H}$, then 
\begin{align}
 \omega(A) = \omega_\varrho(A) = {\rm Tr}(\varrho A) \quad \ \ (A \in {\sf B}(\mathcal{H}))
\end{align}
is the {\it expectation value functional} -- synonymously, {\it state} -- induced on ${\sf B}(\mathcal{H})$ by $\varrho$. Since every local observable algebra ${\sf A}(O)$ is 
contained in ${\sf B}(\mathcal{H})$, any density matrix state $\omega$ as above induces
a -- partial -- state $\omega_{[{\sf A}(O)]}(A) = \omega(A)$ $(A \in {\sf A}(O))$ on 
${\sf A}(O)$. It is not convenient to write the subscript to indicate a partial state and thus we generally won't use it unless ambiguity might arise. 

For the concepts we summarize next, we largely adhere to 
\cite{Wer-Inv,Keyl}. 
A linear, completely positive map $T: {\sf B}(\mathcal{H}) \to {\sf B}(\mathcal{H})$
such that $T({\bf 1}) = {\bf 1}$, where ${\bf 1}$ denotes the unit operator 
in ${\sf B}(\mathcal{H})$, is called a {\it channel}. (Occasionally, to emphasize the property 
$T({\bf 1}) = {\bf 1}$, it is called a non-selective channel.) Here, we are exclusively 
interested in channels of the form 
\begin{align}
  T(A) = \sum_{j = 1}^N V_j A V_j^*\,, \quad V_j \in {\sf B}(\mathcal{H})\,, \ \ 
  \sum_{j = 1}^N V_jV_j^* = {\bf 1}\,.
\end{align}
for any $A \in {\sf B}(\mathcal{H})$, where $N$ is a finite number. A special case is a {\it unitary channel} $T(A) = UAU^*$
with unitary $U \in {\sf B}(\mathcal{H})$. For a causally complete spacetime region $O = (O^\perp)^\perp$, we call a channel {\it localized in} $O$ if the $V_j$ are contained in ${\sf A}(O)$, which entails $T(A) \in {\sf A}(O)$ for all $A \in {\sf A}(O)$ as well as $T(A') = A'$ for all 
$A' \in {\sf A}(O')$ with $O' \subset O^\perp$. (We caution the reader that this is not 
necessarily canonical terminology.) The dual of a channel, $\tau: \omega \mapsto \tau(\omega)(\,.\,) = 
\omega(T(\,.\,))$, is called a (non-selective) {\it operation}; quite generally, an operation maps states
to states under preservation of convex sums. In this paper, we only consider operations which arise as duals of channels, thereby mapping density matrix states to density matrix states. An operation is called {\it unitary} if it is the dual of a unitary channel, and 
it is called {\it localized} in a causally complete spacetime region $O$ if the channel to which it is dual is localized in $O$. Thus, if an operation $\tau$ is localized in $O$, then 
for any $A' \in {\sf A}(O')$, with $O' \subset O^\perp$, and for every density matrix state $\omega$, it holds that $\tau(\omega)(A') = \omega(A')$. 

Now we turn to the situation considered by Sorkin \cite{Sorkin}, in the form presented
in \cite{BosFewRup}. Thus, we consider three spacetime regions $O_{\aaa}$, $O_{\bbb}$ and $O_{\ccc}$ wherein and during which Alice, Bob and Charlie carry out localized operations and measurements on a density matrix state $\omega$ on ${\sf B}(\mathcal{H})$. The regions $O_{\aaa}$ and $O_{\ccc}$ are causally
separated, i.e.\ $O_{\aaa} \subset O_{\ccc}^\perp$, while
the causal future of $O_{\aaa}$ as well as the causal past of $O_{\ccc}$ intersect $O_{\bbb}$.   In fact, for our argument, we need a suffcient amount of 
causal overlap, although in concrete quantum field models, using specific
properties of the quantum field, this could be weakened. In more detail, we take 
$O_{\bbb} = O(r_2)$ together with the regions $O^{(\pm)} \subset O(r_1)$ as described in the previous section. The causal overlap of $O_{\bbb}$ with $O_{\aaa}$ and $O_{\ccc}$ is
assumed to be such that $O^{(-)}$ is contained in $O_{\aaa} \cap O_{\bbb}$, and $O^{(+)}$ is contained in 
$O_{\bbb} \cap O_{\ccc}$ (see Figure 1).
\begin{center}
 \includegraphics[width=16.2cm]{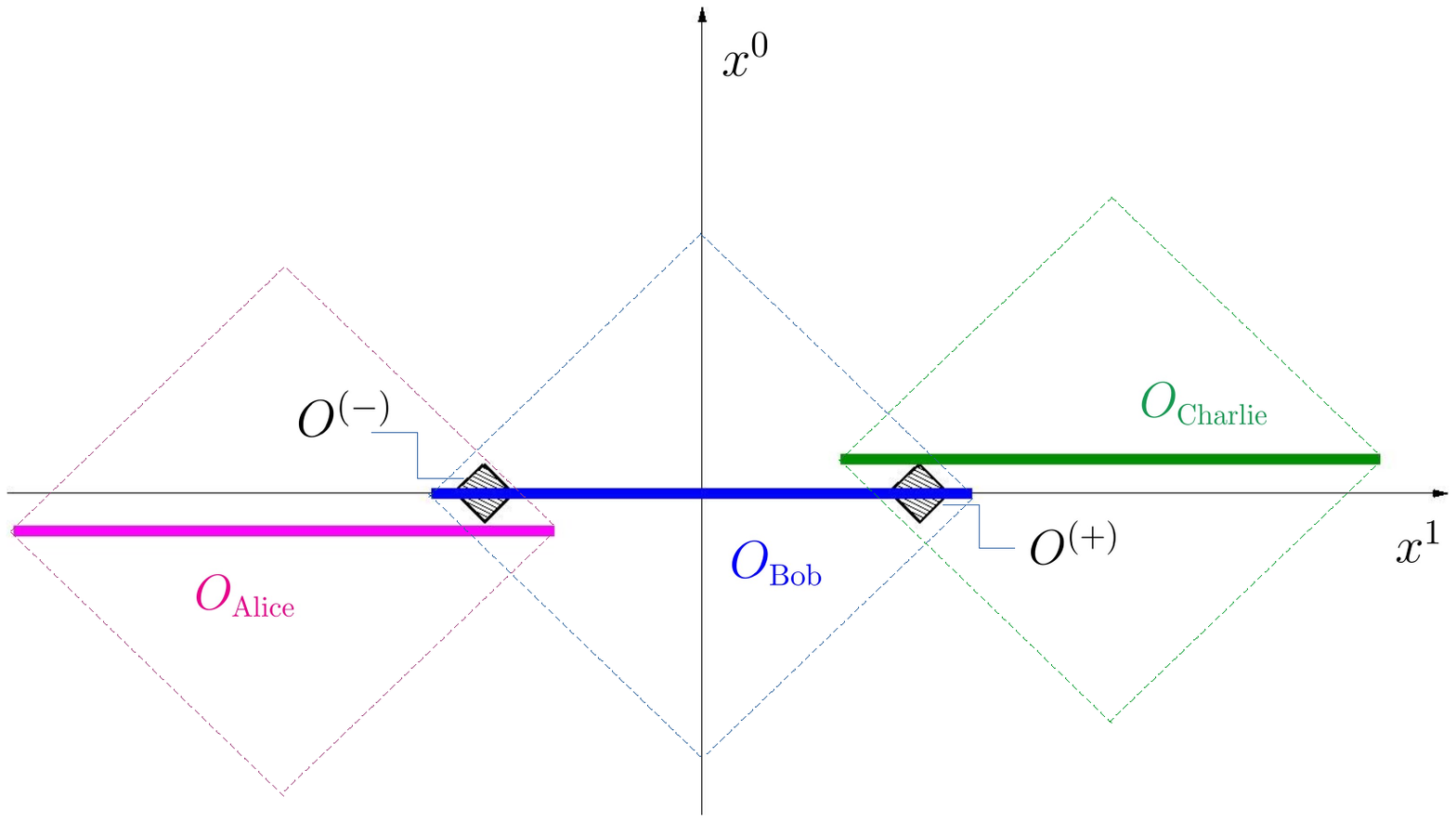}
\end{center}
{\small {\bf Figure 1} The figure depicts the spacetime regions and their relations described in the text.}
\\[8pt]
With this set-up in place, given any density matrix state $\omega$ on ${\sf B}(\mathcal{H})$, we assume that Alice carries out a unitary operation 
$\tau_{\aaa}$ localized in $O^{(-)}$ -- which is contained in $O_{\aaa}$ -- given as 
\begin{align}
 \tau_{\aaa}(\omega)(A) = \omega(W A W^*) 
\end{align}
with some unitary operator $W \in {\sf A}(O^{(-)})$.
If Charlie carries out a measurement by evaluating any (symmetric) operator 
$C$  in  the state $\tau_{\aaa}(\omega)$, the result is 
\begin{align}
 \tau_{\aaa}(\omega)(C) = \omega(WCW^*) = \omega(WW^*C) = \omega(C)
\end{align}
since $W$ is unitary and $O_{\aaa} \in O_{\ccc}^\perp$. This means that Charlie cannot decide, by measurements in $O_{\ccc}$, if Alice has applied the operation $\tau_{\aaa}$ localized in $O_{\aaa}$. However, if Bob carries out operations localized in $O_{\bbb}$, this may change. In particular, assume that Bob carries out the operation 
\begin{align} \label{eq:Q-rot}
 \tau_{\bbb}(\tilde{\omega})(B) = \tilde{\omega}(\check{U}_{\pi,3} B \check{U}_{\pi,3}^*)
\end{align}
on arbitrary density matrix states $\tilde{\omega}$. Since $\check{U}_{3,\pi}$ is a unitary operator in ${\sf A}(O_{\bbb})$, $\tau_{\bbb}$ is a unitary operation localized in $O_{\bbb}$. Thus, for any operator $C \in {\sf A}(O^{(+)})$ -- recall that $O^{(+)}$ is contained in $O_{\bbb} \cap O_{\ccc}$ -- we have 
$\check{U}_{\pi,3} C \check{U}_{\pi,3}^* \in {\sf A}(O^{(-)})$. Consequently, if 
on any density matrix state $\omega$, first Alice carries out operation $\tau_{\aaa}$,
and then Bob carries out operation $\tau_{\bbb}$, then a measurement by Charlie with
$C \in {\sf A}(O^{(+)})$ yields
\begin{align}
 (\tau_{\bbb} \circ \tau_{\aaa} \,\omega)(C) & = (\tau_{\aaa} \,\omega)
 (\check{U}_{\pi,3} C \check{U}_{\pi,3}^*) \\
 & = \omega(W \check{U}_{\pi,3} C\check{U}_{\pi,3}^* W^*)
\end{align}
Since $C \mapsto \check{C} = \check{U}_{\pi,3} C \check{U}_{\pi,3}^*$ maps the von Neumann algebra 
${\sf A}(O^{(+)})$ onto the von Neumann algebra ${\sf A}(O^{(-)})$, Charlie can, by conducting measurements in $O^{(+)}$, determine if Alice has carried out the operation
$\tau_{\aaa}$, once Bob has carried out the ``instantaneous rotation by 180 degrees around the $x^3$-axis'' operation $\tau_{\bbb}$ -- barring the trivial case that $W$ commutes with all operators in 
${\sf A}(O^{(-)})$ -- however, for a proper quantum field theory, the local von Neumann algebras are non-commutative, so there is a rich supply of unitary $W$ and selfadjoint 
$\check{C}$ in ${\sf A}(O^{(-)})$ that don't commute. In other words, even if $\omega$
is the vacuum state, we will in general have many unitary $W \in {\sf A}(O^{(-)})$ and 
selfadjoint $C \in {\sf A}(O^{(+)})$ so that 
\begin{align} \label{eq:not-equal}
 (\tau_{\bbb} \circ \tau_{\aaa} \,\omega)(C) = \omega(W \check{U}_{\pi,3} C\check{U}_{\pi,3}^* W^*) \ne \omega(C) \,.
\end{align}
In fact, such unitary operators $W$ and $C$ are guaranteed to exist whenever 
${\sf A}(O^{(-)})$ is non-commutative. In turn, this is a obviously a consequence of the 
additivity property of the local von Neumann algebras ${\sf A}(O)$ that we have 
formulated above, and the implicit assumption that we have truely a quantum field theory, i.e.\ that ${\sf B}(\mathcal{H})$ is non-commutative. 

We may quickly illustrate the desired non-commutativity of the local algebras, leading to \eqref{eq:not-equal}, by means of a simple example related to the linear scalar Klein-Gordon field (\cite{RS2}). Here, the local von Neumann algebras ${\sf A}(O)$ are generated 
by unitary operators $W(f) = {\rm e}^{i\Phi(f)}$ where the real-valued, smooth test-functions
$f$ have support in $O$. The field operators $\Phi(f)$ are the self-adjoint extensions
of symmetric operators, defined on the Wightman domain (cf.\ \cite{RS2}), fulfilling $\Phi((\Box + m^2)f) = 0$ for some fixed mass term $m \ge 0$, where $\Box$ denotes the d'Alembert operator in Minkowski spacetime. Further properties are 
\begin{align} \label{eq:comm}
 W(f)W(h) = {\rm e}^{-i\mathscr{G}(f,h)/2}W(h)W(f)
\end{align}
for any smooth, compactly supported, real-valued test-functions $f,h$ on Minkowski spacetime $\mathbb{R}^{1 +3}$. Here $\mathscr{G}$ is the causal Green function (or causal propagator) of the Klein-Gordon operator $\Box + m^2$. It arises as 
\begin{align}
 \mathscr{G}(f,h) = \int f(x) (\mathcal{G}h)(x) \, d^4x
\end{align}
with the causal Green operator $\mathcal{G} = \mathcal{G}^+ - \mathcal{G}^-$ mapping real-valued, compactly supported, smooth test-functions $f$ to solutions to the Klein-Gordon equation, i.e.\ $(\Box + m^2)\mathcal{G}f = 0$, such that the Cauchy-data of $\mathcal{G}f$, on any Cauchy-surface, are compactly supported. The causal Green operator is given as the difference of the retarded minus the advanced Green operators, denoted as $\mathcal{G}^\pm$. The vacuum vector $\Omega$ can be characterized through 
\begin{align}
 (\Omega,W(f)\Omega) = {\rm e}^{-w_2(f,f)} 
\end{align}
with the 2-point function
\begin{align}
 w_2(f,h) = (\Phi(f)\Omega,\Phi(h)\Omega) = \frac{1}{2\pi} \int_{\mathbb{R}^3}
  \overline{\hat{f}(\omega_{\bf p},{\bf p})}\hat{h}(\omega_{\bf p},{\bf p})\,\frac{d^3 {\bf p}}{\omega_{\bf p}}
\end{align}
where the hat denotes Fourier transform and $\omega_{\bf p} = \sqrt{|{\bf p}|^2 + m^2}$
\ \,$({\bf p} \in \mathbb{R}^3)$.
The property \eqref{eq:comm} implies, writing $[A,B] = AB- BA$ for the commutator, 
\begin{align}
  \Phi(f) - W(h)\Phi(f)W(h)^* = [\Phi(f),W(h)]W(h)^* = - \mathscr{G}(f,h){\bf 1}
\end{align}
on the Wightman domain, as can be easily checked. Hence, with $\omega(\,.\,) = 
(\Omega,\,.\,\Omega)$ denoting the vacuum state,
\begin{align}
 \omega([\Phi(f),W(h)]W(h)^*) = - \mathscr{G}(f,h) 
\end{align}
and it is not difficult to verify that given any open subset $O$ of Minkowski spacetime, there are smooth, real-valued test-functions $f$ and $h$ having
support in $O$ so that the right-hand side of the last equation is different from
$0$. Then one can replace $\Phi(f)$ by the sequence of bounded symmetric operators 
$T_n = ({\bf 1} + \frac{1}{n}\Phi(f)^2)^{-1}\Phi(f)$ to conclude that for sufficently 
large $n \in \mathbb{N}$, one has 
\begin{align}
 \omega([T_n,W(h)]W(h)^*) \ne 0\,.
\end{align}
Consequently, if we choose especially $O = O^{(-)}$, and set $W = W(h)$ and 
$\check{C} = \check{U}_{\pi,3} C\check{U}_{\pi,3}^* = T_n$, we obtain
\begin{align}
 \omega([\check{C},W]W^*) = \omega(\check{C}) - \omega(W\check{C}W^*) \ne 0\,.
\end{align}
On the other hand, we note that according to the definition of the 
operators $\check{U}_{\pi,3}$, it holds that $\check{U}_{\pi,3}C\check{U}_{\pi,3}^*
= U_{\pi,3}CU_{\pi,3}^*$ for all $C \in {\sf A}(O^{(+)})$ and therefore,
\begin{align}
 \omega(W\check{U}_{\pi,3}C\check{U}_{\pi,3}^*W^*) \ne
\omega( U_{\pi,3}CU_{\pi,3}^*) = \omega(C)
\end{align}
where we used that the vacuum state is invariant under spatial rotations;
$U_{\pi,3}\Omega = \Omega$. We have also used the fact that the quantized scalar Klein-Gordon field in vacuum representation on Minkowski spacetime fulfills all the assumptions that we have listed previously for a quantum field theory, in particular, the split property \cite{Bu-split}.

\section{Superluminal localized state transformations in \\ classical field theory}

\setcounter{equation}{0}

We now wish to demonstrate that similar superluminal localized operations
with the -- geometrical -- significance of ``instantaneous spatial rotations'' are 
also present in classical field theory. To this end, we need a description of 
classical field theory in a local and covariant algebraic setting, in the spirit of the 
approach of Haag and Kastler \cite{HK} for quantum field theory. This has been developed in recent
literature, see \cite{BFRib,Duetsch,Rejzner} and literature cited there. However,
we are mainly focussing on the example of the classical free Klein-Gordon field on 
Minkowski spacetime, so we won't need the theory laid out in the mentioned references
in full generality. Therefore, we present the approach, mostly following
\cite{Duetsch} and \cite{Rejzner}, in a simplified form. 

We start by writing 
$\mathcal{S} = \mathcal{G}(C_0^\infty(\mathbb{R}^{1 +3},\mathbb{R}))$.
This 
is the image of real-valued, smooth, compactly supported test-functions under 
the causal Green operator $\mathcal{G}$ for the Klein-Gordon operator
$\Box + m^2$ on Minkowski spacetime. Thus,
every $\varphi \in \mathcal{S}$ is a smooth, real-valued solution to the Klein-Gordon equation on $\mathbb{R}^{1+3}$, fulfilling $(\Box + m^2)\varphi = 0$. Then
we consider the set of all functions $F: \mathcal{S} \to \mathbb{C}$ which
forms in the usual way a unital, commutative $*$-algebra by defining the algebraic
operations pointwise, i.e.\ $(aF + G)(\varphi) = aF(\varphi) + G(\varphi)$,
$(FG)(\varphi) = F(\varphi)G(\varphi)$, $F^*(\varphi) = \overline{F(\varphi)}$ for
all $\varphi \in \mathcal{S}$ ($a \in \mathbb{C}$, overlining means complex conjugation). The algebra of functions on $\mathcal{S}$ possesses
a unit element, given by ${\bf 1}(\varphi) = 1$. 

In a next step, we define a $*$-subalgebra of the algebra of all functions on 
$\mathcal{S}$, denoted by $\mathscr{P}$. The algebra $\mathscr{P}$ is defined 
to be algebraically generated by the unit element ${\bf 1}$ and all linear functionals 
of the form 
\begin{align}
 F_f(\varphi) = \int_{\mathbb{R}^{1+3}} \varphi(x)f(x)\,d^4x
\end{align}
where $f \in C_0^\infty(\mathbb{R}^{1+3},\mathbb{C})$ is arbitrary. 
(One can enlarge the algebra $\mathscr{P}$ by taking suitable distributional limits of the 
$f$. In the approach presented in \cite{Duetsch,Rejzner}, this is important since it allows e.g.\ to include extended algebra elements
of the form $\tilde{F}(\varphi) = \int_{\mathbb{R}^{1 + 3}} h(x) \varphi(x)^n\,d^4x$.
At this point, however, we won't discuss these matters and refer to the references for further discussion.) Then one can also define local $*$-subalgebras, by defining
for any open subset $O$ of $\mathbb{R}^{1 +3}$,
\begin{align}
 \mathscr{P}(O) = *\text{-subalgebra of} \ \mathscr{P} \ \text{generated by} \
 {\bf 1} \ \text{and all} \ F_f\ \text{with} \ {\rm supp}(f) \subset O\,.
\end{align}
It is obvious that $O_1 \subset O_2 \Rightarrow \mathscr{P}(O_1) \subset \mathscr{P}(O_2)$. Moreover, if $L \in \mathcal{P}_+^\uparrow$, then setting
$\beta_L(F_f)(\varphi) = F_f(\varphi \circ L)$ induces automorphisms of $\mathscr{P}$ such that 
\begin{align} \label{eq:covaction}
 \beta_L(\mathscr{P}(O)) = \mathscr{P}(L(O))\,.
\end{align}
For the functions $\varphi \mapsto P(\varphi)$ in $\mathscr{P}$, one can define
the functional derivative $\delta P/\delta \varphi$ by
\begin{align}
 \left. \frac{d}{ds}\right|_{s = 0} P(\varphi + s \chi) = \int_{\mathbb{R}^{1 + 3}} 
 \frac{\delta P}{\delta \varphi}(\varphi) (x) \chi(x) \,d^4x
\end{align}
where $\varphi$ and $\chi$ are in $\mathcal{S}$. To give some examples, we have 
$\delta {\bf 1}/\delta \varphi = 0$, $\delta F_f/\delta \varphi(\varphi)(x) = f(x)$, 
and for $P(\varphi) = F_f(\varphi)F_h(\varphi)$, we have 
$\delta P/\delta \varphi (\varphi)(x) = f(x)F_h(\varphi) + F_f(\varphi)h(x)$. Note that $x \mapsto \delta P/\delta \varphi(\varphi)(x)$ is a smooth, compactly supported 
function on $\mathbb{R}^{1 + 3}$ which depends (in general, non-linearly) on $\varphi$. With the help of the functional derivative of elements of $\mathscr{P}$,
one can introduce a {\it Poisson bracket} (or, more appropriately, a {\it Peierls bracket}) on $\mathscr{P}$, given by 
\begin{align}
 \{ P_1,P_2\}_{\rm PB}(\varphi) = \int_{\mathbb{R}^{1 +3}} \frac{\delta P_1}{\delta \varphi}
 (\varphi)(x)\, \mathcal{G}\left( \frac{\delta P_2}{\delta \varphi}
 (\varphi) \right)(x) \,d^4x 
\end{align}
for $P_1,P_2 \in \mathscr{P}$. Notice that $\varphi \mapsto \{ P_1,P_2\}_{\rm PB}(\varphi)$ is again in $\mathscr{P}$, and we have the following relations (see \cite{Duetsch}):
\begin{align} \label{eq:relations}
 \{P_1,P_2\}_{\rm PB} = - \{P_2,P_1\}_{\rm PB}\,, \quad \{P_1,P_2P_3\}_{\rm PB} = 
 \{P_1,P_2\}_{\rm PB}P_3 + P_2\{P_1,P_3\}_{\rm PB}
\end{align}
with the algebra product in $\mathscr{P}$, $P_2P_3(\varphi) = P_2(\varphi)P_3(\varphi)$. Additionally, the Poisson bracket also fulfills a Jacobi identity. As a consequence of the causal support properties and the covariance 
of the causal Green operator $\mathcal{G}$ with respect to transformations in $\mathcal{P}_+^\uparrow$ (see e.g.\ 
\cite{Dimock,BenDap} and references cited there), one furthermore obtains 
\begin{align}
 \{P_1,P_2\}_{\rm PB} = 0 \ \ \text{for} \ \  P_j \in \mathscr{P}(O_j) \ \ \text{with} \  \ O_1 \subset O_2^\perp
\end{align}
as well as
\begin{align}
 \{\beta_L(P_1), \beta_L(P_2)\}_{\rm PB} = \beta_L(\{P_1,P_2\}_{\rm PB})
\end{align}
for all $L \in \mathcal{P}_+^\uparrow$ and $P_1,P_2 \in \mathscr{P}$.
\\[6pt]
Hence, we see that the theory of the classical Klein-Gordon field on Min\-kowski spacetime can be formulated in a very similar way as for the quantized field. 
The functions $\varphi \mapsto P(\varphi)$ $(\varphi \in \mathcal{S})$ in $\mathscr{P}$ are 
(simple, polynomial) functions on $\mathcal{S}$, the space of solutions to
the Klein-Gordon equation that have compactly supported Cauchy data. 
This space 
of solutions can be identified with the space of Cauchy data of solutions to the 
Klein-Gordon equation, as we will soon discuss in more detail. The space of Cauchy data naturally corresponds to the phase space for a classical field theory in a Hamiltonian setting, and can be dynamically described with the help of the Poisson bracket (see \cite{WaldQFT} for further discussion). The elements in $\mathscr{P}$
are functions on the phase space and hence, if real-valued, they correspond to 
simple observables for the classical Klein-Gordon field. (As mentioned, the set of observables could be enlarged by taking suitable limits of elements
$P \in \mathscr{P}$.)
Since it is a classical field 
theory, the observable algebra is commutative. In the spirit of \cite{HK}, advocating that in relativistic field theory the observables should be localized and covariant, we also have local algebras $\mathscr{P}(O)$ of observables which can 
be measured within the spacetime regions $O$, and actions of the Poincar\'e transformations by automorphisms with the covariance property \eqref{eq:covaction}. 

Also for the unital $*$-algebra $\mathscr{P}$, states are linear functionals 
$\nu : \mathscr{P} \to \mathbb{C}$ which are positive, $\nu(P^*P) \ge 0$ (and commonly also normalized, $\nu({\bf 1}) = 1$). States may arise through suitable measures $\mu$ on $\mathcal{S}$
(assuming suitable additional structure needed for defining measures has been put in place), as 
integrals 
\begin{align}
 \nu(P) = \int_{\mathcal{S}} P(\varphi)\,d\mu(\varphi)
\end{align}
and for any arbitrarily chosen $\varphi_0$ in $\mathcal{S}$, the Dirac measure 
$\delta_{\varphi_0}(P) = P(\varphi_0)$ is an example. We shall, however, not 
discuss this matter further. 
\\[6pt]
Given any Cauchy-surface $\Sigma$ in $\mathbb{R}^{1 +3}$, with 
future-pointing unit-normal vector field $n^\mu$ along it, the Cauchy-data 
of any $\varphi \in \mathcal{S}$ on $\Sigma$ have compact support, i.e.\
\begin{align} \label{eq:surfacedata}
 u_{\varphi} = \left. \varphi \right|_\Sigma \quad \text{and} \quad v_{\varphi} = \left. n^\mu \frac{\partial}{\partial x^\mu} \varphi \right|_\Sigma
\end{align}
are in $C_0^\infty(\Sigma,\mathbb{R})$. Furthermore, $\mathcal{S}$ carries a 
canonical symplectic form $\sigma$ given by 
\begin{align}
 \sigma(\varphi,\psi) = \int_\Sigma ( u_{\varphi}v_{\psi} - v_{\varphi}u_{\psi})\,d{\rm vol}_\Sigma
\end{align}
where $d{\rm vol}_\Sigma$ denotes the metric-induced volume element on $\Sigma$. It is worth noting that the symplectic form $\sigma$ is independent of the choice of $\Sigma$. For a proof of these properties and further facts that we will use about the symplectic structure of the space of solutions to the Klein-Gordon equation and the relation to the Green operator below, see e.g.\ 
\cite{Dimock,BenDap,WaldQFT} and references cited there. According to the definition
of $\mathcal{S}$, there is for every $\varphi \in \mathcal{S}$ some
$f \in C_0^\infty(\mathbb{R}^{1+3},\mathbb{R})$ so that $\varphi = \mathcal{G}f$. In fact, the map $C_0^\infty(\mathbb{R}^{1+3},\mathbb{R})/{\rm ker}(\mathcal{G}) \to \mathcal{S}$, given by 
$ [f] = f + {\rm ker}(\mathcal{G}) \mapsto \mathcal{G}f$, is a linear bijection, and it 
is also a symplectomorphism upon endowing $C_0^\infty(\mathbb{R}^{1+3},\mathbb{R})/{\rm ker}(\mathcal{G})$ with the symplectic form
\begin{align}
 \kappa([f],[h]) = 
 \int_{\mathbb{R}^{1+3}} f(x) (\mathcal{G}h)(x) \,d^4x\,.
\end{align}

Now let us return to the geometric situation that we have been considering in Figure 1. Our aim is to construct localized rotations on the system of local Poisson algebras $\mathscr{P}(O)$ which preserve the Poisson structure. More precisely,
choosing positive radii $r_1 < r_2$, we have $\overline{B(r_1)} \subset B(r_2)$
for the coordinate balls at $x^0$ defined by \eqref{eq:ball}, and similarly for 
their domains of dependence defined by \eqref{eq:diamond}, $\overline{O(r_1)} \subset O(r_2)$. In the $x^0= 0$ hyperplane which is a copy of $\mathbb{R}^3$, we will introduce for any $0 \le \theta < 2\pi$ a diffeomorphism $\gamma_\theta$ which acts like
a rotation around the $x^3$-axis by an angle of $\theta$ within $B(r_1)$, and 
like the identity outside $B(r_2)$. To this end, we consider the vector field 
$\boldsymbol{f}$ on $\mathbb{R}^3$ given by 
\begin{align}
 \boldsymbol{f} = \eta(r)\left(x^1 \frac{\partial}{\partial x^2} - x^2 \frac{\partial}{\partial x^1}\right)
\end{align}
where $r = \sqrt{ (x^1)^2 + (x^2)^2 + (x^3)^2 }$ is the radius function and 
$\eta : \mathbb{R}_+ \to \mathbb{R}_+$ is a smooth function with 
$\eta(r) = 1$ for $r \le r_1$ and $\eta(r) = 0$ for $r \ge r_2$. Then we take 
$\gamma_\theta : \mathbb{R}^3 \to \mathbb{R}^3$ to be the flow generated by $\boldsymbol{f}$ with flow parameter $\theta$ (so $d \gamma_\theta/d\theta = 
\boldsymbol{f} \circ \gamma_\theta$). It it easy to see that $\gamma_\theta$ 
has the claimed geometric properties. In the next step, we define the linear map
$S_\theta: \mathcal{S} \to \mathcal{S}$ by choosing the Cauchy-surface $\Sigma$ in
\eqref{eq:surfacedata} as the $x^0 = 0$ hyperplane, and setting
\begin{align}
 \left( \begin{array}{c}
         u_{S_\theta \varphi} \\
         v_{S_\theta \varphi}
        \end{array} \right) 
        =
\left( \begin{array}{c} u_\varphi
       \circ (\gamma_\theta)^{-1} \\
        q_\theta v_{\varphi} \circ (\gamma_\theta)^{-1}
        \end{array} \right) 
\end{align}
where $q_\theta = {\rm det}(J_\theta)$, with $J_\theta$ the Jacobian matrix
of $\gamma_\theta$. It is plain to see that, due to the compensating factor 
$q_\theta$, one has 
\begin{align}
 \sigma(S_\theta \varphi, S_\theta \psi) = \sigma(\varphi,\psi)
\end{align}
for all $\varphi,\psi \in \mathcal{S}$, hence $S_\theta$ is a symplectomorphism on the solution space $\mathcal{S}$ for the Klein-Gordon equation with the symplectic form $\sigma$. Note that $q_\theta = 1$ on $B(r_1)$, as well as outside of $B(r_2)$.  

In a further step, we wish to show that the symplectomorphism $S_\theta$ induces
a unit-preserving $*$-algebra morphism $\Upsilon_\theta$ of $\mathscr{P}$ through 
\begin{align}
 (\Upsilon_\theta P)(\varphi) = P(S_\theta^{-1}\varphi)
\end{align}
such that the Poisson bracket is preserved,
\begin{align}
 \{ \Upsilon_\theta (P_1),\Upsilon_\theta (P_2)\}_{\rm PB} =
 \Upsilon_\theta ( \{P_1,P_2\}_{\rm PB})\,.
\end{align}
In the light of the relations \eqref{eq:relations}, it is enough to check the 
preservation of the Poisson bracket for the cases $P_j = F_{f_j}$. To this end,
if $f_j \in C_0^\infty(\mathbb{R}^{1 + 3},\mathbb{R})$, and if $h \in C_0^\infty(\mathbb{R}^{1 + 3},\mathbb{R})$ is chosen with $\varphi = \mathcal{G}h$, then
\begin{align}
 F_{f_j}(\varphi) = \int_{\mathbb{R}^{1+3}} f_j(x) \mathcal{G}h(x)\,d^4x = 
 \kappa([f_j],[h]) = \sigma(\mathcal{G} f_j,\varphi)
\end{align}
Hence, setting $\psi_j = \mathcal{G}f_j$ it follows that 
$\Upsilon_\theta F_{f_j}(\varphi) = F_{f_j}(S_\theta^{-1}\varphi) = \sigma(\psi_j,S_\theta^{-1}\varphi) = \sigma(S_\theta \psi_j,\varphi)$ since $S_\theta$ is a symplectomorphism. On the other hand, we have 
\begin{align}
 \{ F_{f_1},F_{f_2} \}_{\rm PB}(\varphi) = \int_{\mathbb{R}^{1+3}} f_1(x) \mathcal{G}f_2(x) \,d^4x = \kappa([f_1],[f_2]) = \sigma(\psi_1,\psi_2)
\end{align}
from which one can now deduce
\begin{align}
 \{\Upsilon_\theta F_{f_1},\Upsilon_\theta F_{f_2}\}_{\rm PB}(\varphi) = 
 \sigma(S_\theta \psi_1,S_\theta \psi_2) = \sigma(\psi_1,\psi_2) \,.
\end{align}
On the other hand, since $\{F_{f_1},F_{f_2}\}_{\rm PB}(\varphi) = \sigma(\psi_1,\psi_2)$ is independent of $\varphi$ (i.e.\ it is a multiple of the unit element in 
$\mathscr{P}$), we have $\Upsilon_\theta(\{F_{f_1},F_{f_2} \}_{\rm PB}) = 
\{F_{f_1},F_{f_2}\}_{\rm PB}$ and hence
\begin{align}
 \{\Upsilon_\theta F_{f_1},\Upsilon_\theta F_{f_2}\}_{\rm PB} = \Upsilon_\theta(\{F_{f_1},F_{f_2} \}_{\rm PB})
\end{align}
as required so as to show that $\Upsilon_\theta$ is a $*$-algebra morphism of $\mathscr{P}$ preserving the Poisson structure. It is also easy to see from the geometric construction that $\Upsilon_\theta P = P $ for all $P \in \mathscr{P}(\tilde{O})$ with $\tilde{O} \subset O(r_2)^\perp$, and
\begin{align}
 \Upsilon_\theta(\mathscr{P}(O)) = \mathscr{P}(R_{3,\theta}O)
\end{align}
for all $O \subset O(r_1)$, where $R_{3,\theta}$ denotes the space rotation around 
the $x^3$-axis by the angle $\theta$. 

Hence, for the classical Klein-Gordon field on Minkowski spacetime described in the 
algebraic setting in terms of local Poisson algebras, $\Upsilon_\theta$ is a local channel, acting trivially in the causal complement of $O(r_2)  = O_{\bbb}$, and like an ``instantaneous'' space rotation within $O(r_1)$. Thus, in the situation depicted in Figure 1, the operation 
\begin{align}
 \tilde{\tau}_{\bbb} \nu = \nu \circ \Upsilon_{\pi}
\end{align}
on states $\nu$ of $\mathscr{P}$ is the counterpart of $\tau_{\bbb}$ in \eqref{eq:Q-rot} that we had considered before in the quantum field theory framework. Obviously, $\tilde{\tau}_{\bbb}$ is not provided by the action of 
unitary algebra elements, since the algebra $\mathscr{P}$ is commutative. Thus,
whenever $G \in \mathscr{P}$ (or, for that matter, on replacing $\mathscr{P}$ by a suitable extention) fulfills $GG^* = {\bf 1}$, then 
$G P G^* = P$ for all $P \in \mathscr{P}$. 

This said, it should now be clear that there are also channels $\Upsilon_{\aaa}$ for $\mathscr{P}$ which are localized in $O_{\aaa}$ (i.e.\ they act trivially in the causal complement of $O_{\aaa}$) so that for their induced operations $\tilde{\tau}_{\aaa}$
given by $\tilde{\tau}_{\aaa}\nu = \nu \circ \Upsilon_{\aaa}$ we find
\begin{align}
 \tilde{\tau}_{\aaa}\nu(C) = \nu(C)
\end{align}
for all states $\nu$ of $\mathscr{P}$ and all $C \in \mathscr{P}(O_{\ccc})$, while
\begin{align}
 (\tilde{\tau}_{\bbb} \circ \tilde{\tau}_{\aaa} \nu)(C) \ne \nu(C)
\end{align}
for some states $\nu$ and suitable $C \in \mathscr{P}(O^{(+)})$ (cf.\ Figure 1).
For instance, one can choose for $\Upsilon_{\aaa}$ a rotation around some space axis in $O_{\aaa}$, constructed in the same manner as $\Upsilon_{\theta}$ with respect to $O_{\bbb}$. In other words, we have provided, in an algebraic setting for a classical, relativistic field theory, an example of an ``impossible measurement scenario'' where, according to \cite{Sorkin}, the information if Alice has carried out an operation in her lab is mediated by an operation in Bob's lab with ``superluminal speed'' to the lab of Charlie which is causally separated from Alice. 

\section{Discussion}

\setcounter{equation}{0}

We have shown that in the algebraic framework, both in quantum field theory -- under very general assumptions as well as more concretely, for the quantized Klein-Gordon field -- and in classical field theory (for the classical Klein-Gordon field), ``superluminal localized operations'' $\tau_{\bbb}$ occur. They have a geometric significance as ``instantaneous space rotations'' by 180 degrees, and they lead to
the scenario which in \cite{Sorkin} has been connected with the ``impossible measurements scenario'' where (cf.\ Figure 1) Charlie can tell if Alice has carried 
out an operation on a state $\omega$ if Bob carries out $\tau_{\bbb}$ in localized
in $O_{\bbb}$ through the relation
\begin{align}
 (\tau_{\bbb} \circ \tau_{\aaa} \omega)(C) \ne \omega(C)
\end{align}
for some states $\omega$ and some observables $C$ measured by Charlie in $O_{\ccc}$. In this sense, at face value, the ``impossible measurements scenario'' in 
\cite{Sorkin} {\it fails the ping-pong ball test} in the sense that it isn't a feature of quantum field theory only, but also occurs in classical field theory. 

That is not to say, however, that the scenario presented in \cite{Sorkin} was without interest or significance. In fact, various interesting lessons can be learned by having subjected it to our ping-pong ball test. 

First, we see that, as pointed out in \cite{Sorkin} and \cite{AJu,Jubb}, localized operations, both in quantum field theory and in classical field theory, 
are only specified by acting trivially in the causal complement
of the spacetime region wherein they are localized, but they can act superluminally
within that localization region. As we have seen, this includes (unsurprisingly)
``passive'' transformations which are related to (local) symmetries of (the theory of) a physical system. However, carrying them out ``instantaneously'' is actually impossible on kinematical or dynamical grounds. What can really be carried out in a lab on a physical system must be brought about by interaction, and in a relativistic theory, it must respect the principle that ``no action on a system can proceed faster than with the velocity of light'', i.e.\ not leading to superluminal effects. In the local, algebraic setting of quantum field theory, or of classical field theory, one could think of various ways of capturing this principle. A quite strong requirement on operations $\tau$ to be physical could be that they should arise as duals of channels $T$ which obey 
\begin{align}
 T({\sf A}(O)) \subset {\sf A}(J(O))
\end{align}
for {\it all} subsets $O$ of Minkowski spacetime, where $J(O)$ is the 
causal set of $O$, i.e.\ the set of all points which lie on causal curves emanating
from $O$. In \cite{BuVe-U1}, successions of Fermi-Walker transported observables
$t \mapsto \alpha_{L_t}(A)$ for $A \in {\sf A}(O)$ have been considered,
where $\{L_t\}_{t \in I}$, with $I$ a real interval,  is a smooth family of Poincar\'e transformations
such that, for every $x \in O$, $t \mapsto L_t(x)$ is a future-directed, causal curve. One could attempt to restrict the possibility of instantaneous rotations
(or other instantaneous Poincar\'e transformations) in a similar manner. For the discussion of other, related restrictions on local operations in order to prevent them from acting in a superluminal fashion, see e.g.\ \cite{AJu,Jubb}. 

While investigating useful kinematical characterizations of local operations compatible with
the principles of special or general relativity is important -- and may actually remove a gap in the literature on localized operations -- we think that of prime importance is really the aspect that in the lab, the experimenter carries out ``active'' operations, i.e.\ operations that involve interactions with the physical system under consideration. In the framework of quantum field measurement set out in \cite{FVqflm,FewQFM}, the system under consideration, described by a quantum field, interacts with another quantum field, modelling the probe. The interaction is subject to specific conditions on localization and causality which, in consequence, avoid the impossible measurement scenario  for the 
operations resulting from the interaction of system and probe \cite{BosFewRup}. Imposing suitable locality and causality conditions on interactions is also of 
importance in the construction of interacting quantum field theories, see 
\cite{BuFre-Int1,BuFre-Int2} for a recent contribution in this direction, as well
as related discussion in \cite{Duetsch,Rejzner}. 

A second lesson that may be drawn is about the status of the unitary elements $U$
of the local algebras ${\sf A}(O)$ in quantum field theory as operations, or 
more precisely, as giving rise to channels $A \mapsto UAU^*$ inducing local operations. As we have 
mentioned already, this doesn't match too well with how local operations arise
in the algebraic framework of classical field theory because the algebras 
$\mathscr{P}(O)$ are commutative. However, in classical field theory, the 
action of generators of (local) symmetries can be obtained with the help of 
the Poisson bracket and elements $G$ of the (local) Poisson algebras, i.e.\ as 
derivations of the form $P \mapsto \{G,P\}_{\rm PB}$ \cite{Schweber}. Similarly
in quantum field theory, the commutator bracket with (typically unbounded) operators $Q$ affiliated to the local algebras ${\sf A}(O)$ gives rise 
to the derivations $A \mapsto [Q,A]$ generating (local) symmetries. This analogy 
is very familiar when discussing the passage from Hamiltonian mechanics to quantum mechanics, and therefore, in comparison with the classical field theory situation,
the operators affiliated with local algebras should be seen as the generators,
in the commutator bracket, of local channels. The circumstance that in quantum field theory, the
corresponding channels are actually implemented by unitary operators $U$ in the 
local algebras ${\sf A}(O)$, is perhaps more a consequence of the richness of the 
${\sf A}(O)$ and less related to an a priori significance of unitaries in the 
local algebras as implementers of local channels and their associated operations.

\end{document}